# EFFICIENT DETECTION OF SYBIL ATTACK BASED ON CRYPTOGRAPHY IN VANET

Mina Rahbari[1] and Mohammad Ali Jabreil Jamali[2]

[1]Department of Computer Science, Shabestar Branch, Islamic Azad University, Shabestar, Iran
mina_rahbari@yahoo.com
[2] Department of Computer Science, Shabestar Branch, Islamic Azad University, Shabestar, Iran
m_jamali@itrc.ac.ir

## ABSTRACT

*Vehicular communications play a substantial role in providing safety transportation by means of safety message exchange. Researchers have proposed several solutions for securing safety messages. Protocols based on a fixed key infrastructure are more efficient in implementation and maintain stronger security in comparison with dynamic structures. The purpose of this paper present a method based on a fixed key infrastructure for detection impersonation attack, in other words, Sybil attack, in the vehicular ad hoc network. This attack, puts a great impact on performance of the network. The proposed method, using an cryptography mechanism to detection Sybil attack. Finally, using Mat lab simulator the results of this approach are reviewed, This method it has low delay for detection Sybil attack, because most operations are done in Certification Authority, so this proposed schema is a efficient method for detection Sybil attack.*

## 1. INTRODUCTION

Vehicular Ad-Hoc Network (VANET) is a specific type of Mobile Ad-Hoc Network (MANET) that provides communication between (1) nearby vehicles and (2) vehicles and nearby roadside equipments. VANETs are one way to implement Intelligent Transportation System (ITS), a technique for imparting information and communication technology to transport infrastructure and vehicles. It is based on IEEE 802.11p standard for Wireless Access for Vehicular Environment (WAVE). These networks have no fixed infrastructure, and they rely on the themselves for implementing any network functionality. A VANET is a decentralized network as every node performs the functions of both host and router. The main benefit of VANET communication is enhancement of passenger safety by exchanging warning messages between vehicles. VANETs differ from MANETs in high mobility of nodes, large scale of networks, geographically constrained topology, and frequent network fragmentation. Most of the research on VANET is focused on Medium Access Control (MAC) layer and the network layer . VANETS aim to build applications such as collision avoidance, route changing, and so on. Security of vehicular networks is still largely an explored area. VANET, being a wireless network, inherits all the security threats that a wireless system has to deal with. VANET security is critical because a poorly designed VANET is vulnerable to network attacks, and this can compromise the safety of drivers. A security system should ensure that transmission comes from a trusted source and is not a tampered en-route by other sources. It should also strike a balance with privacy because implementing security and privacy together in a system is contradictory. There are various types of possible attacks on VANETs. It is imperative that VANET security should be capable of handling every type of attack. VANET security is different from that of wireless and wired networks because of its unique characteristics of mobility constraints, infrastructure-less framework, and short duration of link between nodes. In a wired network, infrastructure has components for specific functions, for example, routers decide the route to destination while network hosts send and receive messages. Security implementation is relatively easy as networks need to be physically tampered for eavesdropping. Wireless networks use infrared or radio frequency signals to communicate among devices. These networks can be either (a) infrastructure based or (b) infrastructure-less. Infrastructure-based wireless networks are based on Public Switched Telephone Network (PSTN) switches, MSCs, base stations, and mobile hosts. In ad-hoc networks, a type of infrastructure-less wireless networks, nodes perform all operations such as routing, packet forwarding, and network management, and so on. The existing security solutions use traditional digital signature and certificates using Public Key Infrastructure (PKI).

In VANETs, primary focus of security is on safety-related applications. Non safety applications have less stringent security requirements. There is no prior trust relationship between the nodes of VANETs because of its infrastructure-less nature. Any node can join and leave the network at anytime without informing other nodes in vicinity. Cooperative security schemes are more efficient in VANETs as node misbehavior can be detected through collaboration between the number of nodes by assuming that majority of nodes are honest.

In vehicular ad hoc network (VANET), vehicles are exchanged information such as their status, accidental, potentially dangerous situations and … In the form of messages between each other. With interpreting and processing these messages, drivers become aware of the situation and appropriate decisions are taken to prevent accidents. Obviously, the publication of false information in addition to reduce network performance lead to financial and even physical damage. Sybil attack, is a serious threat as it impairs the functionality of VANETs. In this attack, an attacker node sends messages with multiple identities to other nodes in the network. The attacker simulates several nodes in the network. The node spoofing the identities of other nodes is called malicious node/Sybil attacker, and the nodes whose identities are spoofed are called Sybil nodes. Almost every other attack can be launched in a network in the presence of Sybil attack. One possibility could be an illusion of a traffic jam or accident so that other vehicles change their routing path or leave the road for the benefit of the attacker. Sybil attacker can also inject false information in the networks via some fabricated nonexistent nodes [1 , 2 , 3].

Using a node with nature of Sybil, may are affected on some types of network services such as routing, data traffic congestion in the network, fair allocation of resources, make decisions, recognition abuse any may reduce performance and quality of services in these networks. purpose of this paper present a method for detection impersonation attack based on cryptography.

## 2. ATTACKS ON VEHICULAR NETWORKS

Before designing any security solution for VANETs [17,18], we should know different types of security threats, their capabilities, and the types of attackers also.

### 2.1. Classification of Attackers

Attackers can be classified according to scope, nature, and behavior of attacks [19,20]. Some types of attackers are discussed in following paragraph:

1. Some attackers eavesdrop only on the wireless channel to collect traffic information which may be passed onto other attackers. As these attackers do not participate in the communication process of the network, they are called passive attackers. On the other hand, some attackers either generate packets containing wrong information or do not forward the received packets.

These are called active attackers.

2. Attacker may be an authentic member of a VANET having authentic public keys and access to other members of the network. Such attackers are called insider. Outside attackers (outsider) are intruders and they can launch attacks of less diversity.

3. Some attackers are not personally benefited from the attack. Their aim is to harm other members of the network or disrupt the functionality of a VANET. These attackers are malicious.

On the other hand, rational attacker seeks personal benefit and is more predictable in terms of type and target of the attack.

4. Local attacker launches an attack with a limited scope, that is, an attack is restricted to a particular area. An attack can be extended, where an attacker can control several entities distributed across the network.

## 2.2. Types of Attacks

Owing to the large number of autonomous network members and the presence of human factor, misbehavior of nodes in future vehicular networks cannot be ruled out. Several types of attacks [20] have been identified and classified on the basis of layers used by the attacker. At the physical and link layer, an attacker can disturb the network system by overloading the communication channel with useless messages. An attacker can inject false messages or rebroadcast an old message also. Some attackers can tamper with an OBU or destroy an RSU. At network layer, an attacker can insert false routing messages or overload the system with routing information. Privacy of drivers can be disclosed by revealing and tracking the position of drivers. Some of these attacks are briefly explained subsequently.

### 2.2.1. Bogus Information

In this case, attackers are insiders, rational, and active. They can send wrong information in the network so that it can affect the behavior of other drivers. For example, an adversary can inject wrong information about a nonexistent traffic jam or an accident diverting vehicles to other routes and freeing a route for itself.

### 2.2.2. Cheating with Sensor Information

This attack is launched by an attacker who is insider, rational, and active. He uses this attack to alter the perceived position, speed, and direction of other nodes in order to escape liability in case of any mishap.

### 2.2.3. ID Disclosure

An attacker is insider, passive, and malicious. It can monitor trajectories of a target vehicle and can use this information for determining the ID of a vehicle.

### 2.2.4. Denial of Service (DOS)

Attacker is malicious, active, and local in this case. Attacker may want to bring down the network by sending unnecessary messages on the channel. Example of this attack includes channel jamming and injection of dummy messages.

### 2.2.5. Replaying and Dropping Packets

An attacker may drop legitimate packets. For example, an attacker can drop all the alert messages meant for warning vehicles proceeding toward the accident location. Similarly, an attacker can replay the packets after that event has been occurred to create the illusion of accident.

**2.2.6. Hidden Vehicle**

This type of attack is possible in a scenario where vehicles smartly try to reduce the congestion on the wireless channel. For example, a vehicle has sent a warning message to its neighbours and it is awaiting a response. After receiving a response, the vehicle realizes that its neighbour is in a better position to forward the warning message and stops sending this message to other nodes. This is because it assumes that its neighbour will forward the message to other nodes. If this neighbor node is an attacker, it can be fatal for the system.

**2.2.7. Worm Hole Attack**

It is challenging to detect and prevent this attack. A malicious node can record packets at one location in the network and tunnel them to other location through a private network shared with malicious nodes. Severity of the attack increases if the malicious node sends only control messages through the tunnel and not data packets.

**2.2.8. Sybil Attack**

In this attack, a vehicle forges the identities of multiple vehicles. These identities can be used to play any type of attack in the system. These false identities also create an illusion that there are additional vehicles on the road. Consequence of this attack is that every type of attack can be played after spoofing the positions or identities of other nodes in the network.

# 3. DETECTION OF SYBIL ATTACK

In literature, different techniques are proposed for detection of Sybil attack in VANETs. Sybil attacks are always possible in the absence of any logical centralized authority. As there is no centralized entity in VANETs, detection of Sybil attacks is very difficult. Some constraints such as validating all entities simultaneously by all nodes and strict coordination among entities are necessary for detection of a Sybil attack. Some techniques are described below.

## 3.1. Directional Antenna

This technique, can be used to detect Sybil attack discussed in[3]. This method is used to direction of arrival packets and It checks whether the messages has been come from forged neighbours or neighbours real. This method is not perfect because it sometimes does not detect some attacks.

## 3.2. Propagation Model

Sybil attack can also be detected by using a propagation model as described in[4,5,6]. In this technique, the received signal power from a sending node is matched with its claimed position. By using this method, received signal power can be used to calculate the position of the node. If both the positions (calculated and claimed) do not match, this may be a Sybil node. This technique is unsuitable for detection of a Sybil attack as a malicious node can use the same propagation model to compute the transmission signal strength required to fool detection system in estimating the next position of the node. Signal strength approach has a limited accuracy. Small-scale attacks cannot be detected. It is very difficult for a malicious vehicle to change signal strength distribution. Any change in signal strength will, therefore, be detected by a receiver. If each vehicle is given limited space, malicious vehicles can fabricate only few Sybil nodes. More realistic radio propagation model is required to support high mobility of nodes in VANETs.

### 3.3. Resource Testing

This technique, can be used to detect Sybil attack discussed in[4,6,7,15]. It is assumed that every physical entity is equipped with limited computational resources. A typical puzzle is given to all the nodes in the network for testing computational resources. If resources of a single node are used to simulate multiple entities, any particular entity will be resource constrained in computation, storage, and bandwidth. This approach is not suitable as an attacker may have more computational resources when compared with honest nodes. Yet another problem is that this technique may create network congestion because more number of requests/replies are used for identification of nodes. Radio resource testing can also be used for detecting Sybil nodes. It is based on the assumption that any node has only one radio so any radio cannot send and receive more than one channel at a time. This technique also fails because the attacker can use multiple radio devices simultaneously.

### 3.4. Detection and localization of nodes

This technique, can be used to detect Sybil attack discussed in [5,8]. This method is based on finding the  physical location  of nodes  and comparing  it  with the  vehicle's position is to claim .  so this attack is discovered .  This solution is  the  geometric method and also uses data obtained from GPS.

### 3.5. Public Key Cryptography

Security issue of Sybil attacks can be solved by using public key cryptography and authentication mechanism as described in [10,11]. In this security solution, signatures are combined with digital certificates and asymmetric cryptography is used. Certificates are issued by CA and there is a hierarchy of these CAs. For each region, there is one CA. These CAs communicate with each other through secure channel and keep track of issued certificates used by every signed message. This technique can prevent Sybil attacks as only messages with valid certificates are considered and invalid messages are ignored. The only requirement is that each node should be assigned one certificate at a time. For privacy implementation, these certificates are changed from time-to-time. But in VANETs, it is difficult to deploy PKI as there is no guarantee of the presence of infrastructure. It is very complex, consumes large memory, and time consuming as well.

### 3.6. Timestamp Series

This technique, can be used to detect Sybil attack discussed in [9] In this approach, proposed a timestamp series approach to defend against Sybil attack in a vehicular ad hoc network (VANET) based on roadside unit support. it discover that it would be rare for arbitrary two vehicles to pass through a few different RSUs (far apart from each other) always at the same time. Therefore, if a traffic message sent out by any vehicle contains several timestamps issued to this vehicle by the previously passed RSUs, Sybil attack can be detected if multiple traffic messages contain very similar series of timestamps. This method has challenges, for example If RSUs are located at intersections, it may make the Sybil attack detection difficult, so this method not suitable approach to detect Sybil attack.

## 4. THE PROPOSED SCHEMA

The  proposed schema  uses encryption mechanism to  detects attack and provides  Four  security aspects are concerned in this method:

### 4.1. Authentication

Every receiver vehicle should make sure of message transmitter's authority and authenticate it. In normal network system, each vehicle must have a certificate for transmission, and this allows each vehicle to transmit even if it considered as adversary, common idea is the use of Certificate Revocation List (CRL), CRL will keep the ability for the vehicle to transmit, if any vehicle receives information from a revoked vehicle it will accept the information and apply the id of the sender to the CRL, if the id in the list the receiver will ignore the message, otherwise it will take it, this procedure causes network overhead for frequent retransmission of CRL and causes high computation overhead for each vehicle when receiving any information, and again allows the adversary vehicle to transmit, in some situations the receiving vehicle may accept the information received from adversary, as not all vehicles have the updated CRL. In this work provide each vehicle with special certificate; this certificate will insure the intention status of the vehicle, a Valid Certificate (VC) will be given to the valid vehicle (I mean: not adversary), and Adversary Certificate (AC) for adversary vehicle. Use of this idea cause to performance is increased. Once, vehicle requested for key or send a message, if it has a Valid Certificate key Be assigned to him else if it has a Adversary Certificate, not Be assigned to him and Will not receive a message from it[10,11,14].

## 4.2. Non-repudiation

Every vehicle should put part of its personal information so it can be recognized in the case of crime occurrence and insurance. Thus, repudiation becomes impossible by the transmitter. vehicle's identity should be attached to the message, so it can be tracked whenever desired and Non-repudiation established. Accordingly, vehicle tracking is only allowed just for authorized organization. So the vehicle should encrypt its identity and only authorized organizations is capable of decryption. Hence, encryption of car's identity should be done by means of assigned public key (PU) from authorized organization and be put in a distinct field in original message. Since vehicle's identity is encrypted by a public key, other vehicles are not able to recognize it and just authorized organization own the private key associated to the public key can access its identity.

## 4.3. Privacy

Personal information of vehicles and drivers shouldn't be accessible by other vehicles and the anonymity should be preserved to stop tracking. The exception is for authorized organizations. in this work we use private key to provided privacy.

## 4.4. Data Integrity

The transmitted message should contain valid information not to be altered by attackers. the hash function in addition to encryption of messages ,it summarizes them, and so enhance the network performance. Other methods of encryption are time consuming and reduce network performance. so We to raise the efficiency use the hash function to encryption. Hash message authentication codes (HMAC) sign packets to verify that the information received is exactly the same as the information sent. This is called integrity. HMACs provide integrity through a keyed hash, the result of a mathematical calculation on a message using a hash function combined with a shared, secret key.

This method is easier to manage the CA, a city or a country are divided into the different regions(n). It's noticeable that one key pairs of authentication key and public key are needed for message exchange in each region, thus regions under CA coverage should not have any common area [7,8,10].

$$R = \{R_1, R_2, ..., R_n\}$$

$$R_i \cap R_j = \phi \qquad 1 \le i \quad , \quad j \le n \quad , \quad i \ne j$$

$CA_l$, manages a local public key Which is valid only in its areas. $CA_h$, is hold the initial certification and information of vehicles.

## 4.5. How It Works

Sybil attack detection phases is shown in Figure 1. in this figure malicious node with M, Sybil node with S and The Node that is the identity of his spoofing with A is labaled.

In phase1, each vehicle should be registered in a group and receive its public authentication key (AK) before any message transmission. For signing a message, the vehicle uses group authentication key and encryption function and sends it along with original message to other vehicle and RSU. Therefore it is not obligatory for each member to have other members' private information such as their identity and public key for authenticating them. Receivers verify a member's authenticity by signature verification. It's attained by reconfirmation of encryption function with authentication key to the received message and comparing the result to the signature. Also, receivers can make sure of transmitted data integrity [16].

$$\{M, H_{AK}(M), CA_h, OBU_{ID}\}$$

$OBU_{ID}$, is as follows:

$$OBU_{ID} = \{ID_A, H_{SKA}(ID_A \mid H_{AK}(M))\}_{CA_l}$$

In phase2, Because RSU don't have a private key of $CA_l$, so RSU can not decrypt the message. it sending a request to $CA_l$ to decrypted of the $OBU_{ID}$, in this phase decrypted only $ID_A$.

In phase3, because $CA_l$ don't have private key of vehicle A, so $CA_l$ can not decrypted $HSK(ID_A|H_{AK}(M))$ , therefore send a request private key of vehicle A to the $CA_h$.

In phase4, $CA_h$ reply private key of vehicle A to $CA_l$ and $CA_l$ attained by reconfirmation of encryption function with key of vehicle A to the $(ID_A|H_{AK}(M))$ and comparing the result to the $H_{SK}(ID_A|H_{AK}(M))$. Also, $CA_l$ can detect the Sybil attack , if result of this comparison is different. The Notations used during a Message transmission in 4 phase are shown in Tab.1.

Algorithm used for Sybil attack detection

1- $EH(PU_{AK}(M))$ from source node S
2- $EH(SK_A(ID_A| H_{AK}(M)))$ from source node S
3- $E(PU_{CA}(ID_A, HSK_A(ID_A| H_{AK}(M))))$ from node S
4- $SEND(RQST(M, H_{AK}(M), CA_h, OBU_{Id})$ from source node S to local RSU other vehicle in local region)
5- $EH(PU_{AK}(M))$ in RSU and $IF(H_{AK}(M)==H_{AK}(M))$ THEN go to step 7 else go to step 6
6- REPORT to $CA_l$ "the message is fault"
7- $D(SK_{Cl}(ID_A, HSK_A(Id_A| H_{AK}(M)))$ in $CA_l$
8- $REQST(PU_A)$ to $CA_h$
9- $RPLY(PU_A)$ to $CA_l$
10- $EH(SK_A(ID_A|H_{AK}(M)))$ and $IF(HSK_A(ID_A|H_{AK}(M))==HSK_A(ID_A|H_{AK}(M)))$ THEN Sybil attack detect.

Table 1.  Notation

| Notation | Means |
|----------|-------|
| RQST | Request from source node |
| RPLY | Reply from VANET server |
| SEND | Send key from VANET server to destination |
| E (…) | Encryption of Message |
| EH(…) | Encryption of Message with hashing function |
| D (…) | Decryption of Message |
| $PU_A$ | Public key for source node A |
| $SK_A$ | Private Key for source node A |
| M | Original message |
| AK | Shared key between all nodes is located in a area |
| $H_{AK}(M)$ | Encryption message with Key AK |
| $CA_h$ | Home CA or initial CA |
| $CA_l$ | Local CA |
| $ID_A$ | Identifier of vehicle |

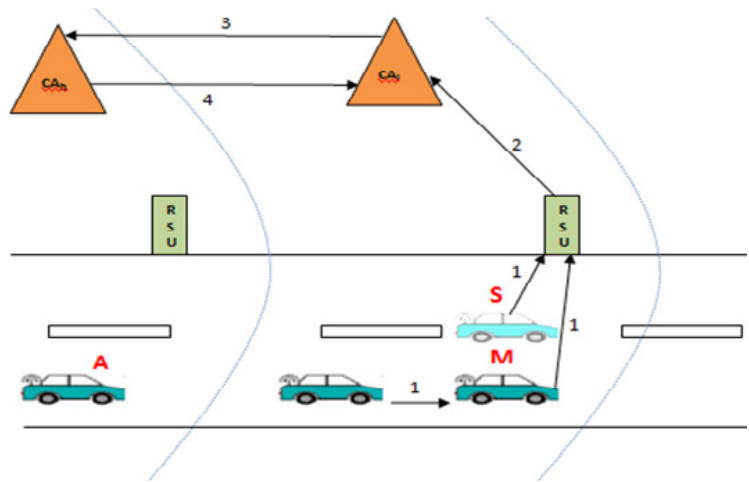

Figure 1.  Phases of detection of Sybil attack

## 4.6. Analysis and Evaluation

A good security mechanism has short delay for encryption, decryption and key exchange. In this proposed scheme is used Matlab tool for simulation. for sending safety message vehicles generate message digest by means of HMAC function and encrypt ID with P-224 curve, in this simulation number of messages that report accident is assume is 5 and number of vehicle is variable to draw these diagrams.

The HMAC operation is very faster than encryption and its delay is not considerable in comparison with encryption delay. In reception of message, the vehicle only generate message digest with AK and compares it with received message digest that takes very short time. Other decryption processes are accomplished by CA that does not influence overall delay. Since frequency of safety message reception is more than its transmission, this method is acceptable.

According above mentioned process, a total delay (D) occurs that this delay is related to factors such as degree of closeness to RSU, connection style of components and overload of the components (RSU, CA).

Execution time of this algorithm is low, because most operations are done in Certification Authority, so the proposed method is a best method for detection of Sybil attacks.

Calculate of total delay:

$$\text{Total delay} = T_1 + T_2 + T_3 + T_4 \qquad (1)$$

$T_1$ means, delay of phase$_1$(delay of broad cast the message to other vehicle and RSU ) and $T_2$ means, delay of phase$_2$(delay of actions in RSU, such as receive messages and categorize messages based on type of event and management of key and compare message with hashed message to find message encrypt with valid key) and $T_3$ means delay of phase$_3$(encryption of messages in CA$_L$) and T4 means delay of phase$_4$(delay of receive key from CA$_h$).

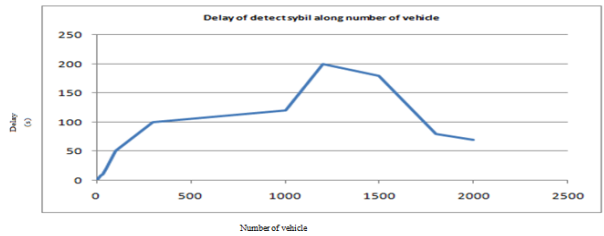

Figure 2. Delay of detect Sybil along number of vehicle

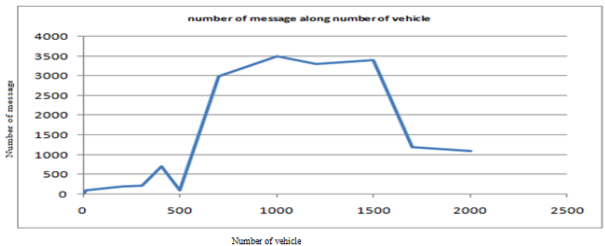

Figure 3. Number of message along number of vehicle

## 5. CONCLUSIONS

This paper presents a method base on cryptography to detect Sybil attack in VANET. Result of simulation shown that Execution time of this algorithm is low, because most operations are done in Certification Authority, so the proposed method is a best method for detection of Sybil attacks. The simulations indicates that, delay of detection Sybil attack depends on the number of messages not to number of vehicles. In our future work we would like to discover location of malicious node, because this nodes is important problem in this type of attacks, prevents of other attacks if malicious nodes is identify. This proposed schema have a problem, that, if nodes move to other rejoins, detection of Sybil attack does not work properly, so in future work we would like to improved this method to detection of Sybil attack execute properly and completely.

**Authors**

**Mina Rahbari**   is currently a researcher at security in Ad hoc network, she is student of master soft engineering in azad Islamic university of shabestar, iran.

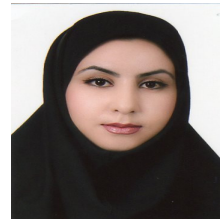